\documentclass{aa}
\usepackage{epsfig}
\begin{document}
\title{Thermal structure of gas in relaxed clusters of galaxies}
\author{E. Fischer}
\institute{Auf der Hoehe 82, 52223 Stolberg, Germany}

\date{Received 2.1.2004 / Accepted }

\abstract
{Gas clouds under the influence of gravitation in thermodynamic equilibrium
cannot be isothermal due to the Dufour effect, the energy flux induced by
density gradients. In galaxy clusters this effect may be responsible for most
of the "cooling flows" instead of radiative cooling of the gas. Observations
from XMM-Newton and Chandra satellites agree well with model calculations,
assuming equilibrium between energy flux induced by temperature and density
gradients, but without radiation loss.}

\maketitle


\section{Introduction}
It has been known for long years that in most clusters of galaxies the
intracluster plasma is far from being isothermal, exhibiting strong decrease
of the plasma temperature in the vicinity of individual large galaxies. This
effect is commonly attributed to "cooling flows", that means, increasing
radiative energy loss with increasing density of the gas, which is
gravitationally attracted by the galaxies. But recent high resolution data
from the XMM-Newton and Chandra satellites have shown that this cooling flow
model does not fit to the observations. This has newly kindled a vivid
discussion on the the origin of the discrepancy.

Already in former years the fact was addressed that thermal conduction would
wash out the temperature gradients, if the thermal conductivity of the plasma
were of the magnitude given in the classical book by Spitzer
(\cite{spitzer}). As a possible explanation it has been proposed that
magnetic fields may reduce the conduction. Several authors claim very
different reduction factors of the thermal conductivity (see e.g.Voigt et al.
\cite{fab}, Markevitch et al. \cite{marke}) to fit the data. The existence of
magnetic fields in the order of $0.1 - 1\rm{\mu G}$ has been confirmed by
various observations (see e.g. Dolag et al. \cite{dolag}). But to produce the
required reduction, ordered magnetic fields of a still higher magnitude would
be necessary, as the heat flux reduction works only perpendicular to the
magnetic field lines and there is not yet a generally agreed model to decide,
to which extent magnetic fields can influence thermal conductivity. Narayan
and Medvedev \cite{naray} have shown that fields fluctuating on a large range
of length scales can produce only minor changes in the thermal conductivity,
while others (see Chandran and Maron \cite{chan} and references therein) give
reduction factors as high as 10.

One problem with all these reduction factors is, that they would affect the
temperature profiles preferably at high temperatures in the outer boundary of
the "cooling flow" region, while radiative cooling is most effective near the
centre. But observations exhibit just the opposite effect (Kaastra et al.
\cite{kaastra})

In all these discussions it is implicitly assumed that in thermodynamic
equilibrium conduction leads to a uniform temperature, though it is well
known that density gradients may also cause a flux of energy and thus induce
temperature gradients, an effect known from the textbooks as "Dufour effect"
(see e.g. Hirschfelder et al. \cite{hirsch}). Under terrestrial conditions
this effect is regarded mostly as a second order correction to "normal"
conduction, that means, to energy flux due to temperature gradients. We are
more familiar with the "reciprocal" effect, the diffusion of particles caused
by temperature gradients, known as thermal diffusion. But under the condition
of cosmic plasmas, where stationary density gradients are maintained by
gravitational fields, the Dufour effect can considerably alter the
equilibrium conditions. Even in relaxed systems there will be stationary
temperature gradients to balance the energy flux caused by density gradients.

Normally in a gas we associate "equilibrium" with constant pressure, density
and temperature. But in a system, which is influenced by volume forces such
as gravitation, the equilibrium state may well exhibit gradients of the state
variables. Mass flow equilibrium in these systems is obtained by the balance
between the gravitational force and the counteracting pressure gradient,
which is related to the density gradient by some equation of state.

But if there is a density gradient, energy equilibrium can be obtained only,
if there is also a temperature gradient, balancing the Dufour effect, to
obtain zero net energy flux. Below we will derive the equilibrium conditions
for a system, which is fully relaxed with respect to flux of mass, momentum
and energy, within the scope of a simplified kinetic model.

In the next section we will discuss the equilibrium conditions in cluster
plasmas. In the third section a simplified kinetic model is developed to
determine the essential consequences of the Dufour effect. In section 4 the
results are compared with observational data from XMM-Newton and Chandra, and
in section 5 conclusions are drawn with respect to the "cooling flow" region
as well as for the equilibrium conditions of complete galaxy clusters.

\section{Equilibrium relations}
Speaking of equilibrium means that the net fluxes of all properties
determining the state of a system are locally balanced. Complete local
equilibrium will of course never be established in a system with radiation
losses or with irreversible chemical reactions. But as long as these are not
the dominant exchange processes, quasi-equilibrium conditions can be
formulated, considering only the dominant transport processes.

To identify the dominant transport processes, it is convenient to compare the
time constants for the local change of temperature, density or composition
due to some loss and production process or flux. Common to all discussions on
the energy balance of the "cooling flow" region of intracluster gas is, that
radiation is regarded as the only energy loss process. This loss has to be
balanced by energy gain from inflowing hot matter, by uptake of gravitational
energy from inflowing gas, by heat conduction from outer hot regions and
energy production by supernovae in galaxies or active galactic nuclei.

The Dufour effect adds a further transport process to the balance, which acts
like an additional cooling, opposite to the heating processes mentioned
above. It can be regarded as a thermal sedimentation of low energy particles
into the vicinity of a central gravitating mass. Microscopically it is
closely connected to the "normal" thermal conduction, as both result from the
same process: the transport of kinetic energy of the plasma particles in an
inhomogeneous medium. If there is thermal conduction, there is always one
fraction caused by temperature gradients and one caused by density gradients.
The relation between the two parts will be discussed more in detail in the
next section.

To find out, which of the heating or cooling processes preferentially
contributes to the balance we compare the time constants of the corresponding
processes. In the "classical" cooling flow model only two processes are
considered: radiation losses and inward flow of hot matter. The time constant
for radiative cooling is $\tau _r= (3nkT)/(\Lambda(T)\,n^2)$, where $\Lambda
(T)$ is the cooling function (see e.g.Voigt et al. \cite{fab}). In the
temperature and density range observed in the centre of galaxy clusters
typical values are $kT=3\,\rm{keV}$ and $n=10^{-2}\rm{cm}^{-3}$, resulting in
a cooling time $\tau _r = 1.2\times 10^{17} \rm{sec}$. As this is much longer
than the collapse time $\tau _c = 1/\sqrt{\rho G}= 3\times 10^{16}\,
\rm{sec}$, equilibrium can always be reached between gas inflow and radiation
loss.

There are, however, two mayor problems with this model. It does not fit to
the observations, as the increasing cooling rate should increase with
increasing density, so that the temperature profile should exhibit a sharp
dip in the centre, which is not observed. Secondly it neglects thermal
conduction, which has time constants much shorter than radiation, so that it
could wash out all gradients established by radiative cooling. If thermal
conduction were at the theoretical value given by Spitzer \cite{spitzer}
$\lambda =4.6\times 10^{-7}\, T^{5/2} \rm{erg \;cm^{-2}K^{-1}sec^{-1}\,}$,
for a cooling flow region of $r_c =50 \;\rm{kpc}$ the conduction time scale
is $\tau _{\lambda} =nkr_c^2 /\lambda = 10^{16}\,\rm{sec}$, much less than
the cooling time scale. Even if thermal conduction is reduced by the presence
of magnetic fields, as proposed by several authors (see Narayan and Medvedev
\cite{naray}, Chandran and Maron \cite{chan}), it remains a dominant
transport process.

But as has been mentioned above, this does not imply that the temperature
should be constant. Due to the opposite effects of temperature and density
gradients also under the condition of local thermal equilibrium the
temperature will vary in radial direction. In the next section we will derive
this equilibrium condition under the assumption that conduction is dominant
and radiation losses can be neglected.

\section{Kinetic model}
To determine the equilibrium conditions in a plasma under the influence of an
outer force field such as gravitation, we determine the net flux of mass and
energy through a control area $\Delta F$ under the condition that the
particles in the gas have a Maxwellian distribution everywhere, but with
number density and mean energy varying perpendicular to the control area. The
net flux is calculated then under the assumption, that any particle crossing
the control area retains its velocity and direction, which it has obtained in
the last collision, one mean free path from the surface, and in addition is
subjected to some acceleration from the force field.

In the energy balance the gain and loss of kinetic energy due to the changing
gravitational potential will be omitted, as it cancels out from the balance
exactly, when the net particle flux through the control area is zero. As we
are interested only in the equilibrium condition, the absolute magnitude of
the energy or particle flux is not important, as long as it dominates over
other transport, production or loss processes, which appears to be justified
for most of the "cooling flow" region in clusters.

Denoting the direction perpendicular to the surface as z and the angle
between this and the direction of particle motion as $\theta$, the flux of
some property $G$ through the surface is
\begin {equation}
\label{jj}
\textstyle{j=\int \Delta F \cos \theta (j^+ - j^-) d \Omega,}
\end {equation}
where $j^+$ and $j^-$ are the normal components of the flux of $G$ carried by
particles moving towards the surface in positive and negative $z$ direction
under the angle $\theta$. The integration has to be taken over half the solid
angle $\pi /2\ge \theta\ge 0$. Denoting the amount of property $G$ carried by
particles moving with velocity $v$ by $g(v)$, the flux components of $g(v)$
are
\begin{equation}
\label{j+}
j^+=(g(v)-\frac {dg(v)}{dz}\lambda \cos \theta )(v \cos \theta + b \frac
{\lambda}{v})
\end{equation}
\begin{equation}
\label{j-}
j^-=(g(v)+\frac {dg(v)}{dz}\lambda \cos \theta )(v \cos \theta - b \frac
{\lambda}{v})
\end{equation}
$\lambda$ is the mean free path, which may depend also on the velocity. The
last term denotes the acceleration of the particles during their flight from
the last collision to the control surface. Integrating eq.(\ref{jj}) over the
solid angle yields the equation
\begin{equation}
\label{j}
j= -\pi \Delta F \left ( \lambda v
\frac{dg(v)}{dz}  -\frac {2b\lambda}{v} g(v)
\right )
\end{equation}

To determine mass and energy transport by particles moving with velocity $v$,
$g$ has to be set to
\begin{equation}
\label{gM}
g_{\rm M}= n(z) m f(v,z) dv, \hspace{.6cm} g_{\rm E}=n(z) m \frac{v^2}{2}
f(v,z) dv
\end{equation}
In case of a neutral gas $n(z)$ and $m$ are the numerical density and mass of
atoms, $f(v,z)$ is the distribution function
\begin{equation}
\label{f}
f(v) dv=\frac{4}{\sqrt{\pi}}\frac{v ^2}{\beta ^3}e^{-\frac{v ^2}{\beta ^2}}
dv
\end{equation}
$\beta (z)$ being the abbreviation $\beta (z)=\sqrt {2kT(z)/m}$. To calculate
the quantity $dg/dz$ we also need the derivative
\begin{equation}
\label{fz}
\frac{df}{dz} = \frac{d\beta}{dz}
\frac{4}{\sqrt{\pi}}\left(\frac{2v ^4}{\beta ^6}\!-\!\frac{3v ^2}{\beta ^4}\right
)e^{-\frac{v ^2}{\beta ^2}} = \frac{1}{\beta}\frac{d\beta}{dz}\!\left
(\frac{2v ^2}{\beta ^2}\!-3\!\right)f(v)
\end{equation}

While in a neutral gas the parameter $b$ in eq.(\ref{j}) is the normal
gravitational acceleration, in a plasma the action of the force is somewhat
indirect. Energy transport is mediated preferably by electrons. Thus in
eqs.(\ref{gM}) $n(z)$ and $m$ are the electron density and mass. But in this
case acceleration by the external force is acting also on the ions, which
transfer the force to the electrons by Coulomb interaction. Thus the
parameter $b$ in eqs.(\ref{j+}) and (\ref{j-}) is not the gravitational
acceleration of free moving electrons. But for the equilibrium condition of
zero transport the absolute magnitude of $b$ is not relevant, when we only
want to know, under which conditions the net flux of electrons and the flux
of energy carried by them are in balance.

An additional difference between neutral gas and plasma results from the fact
that the collision cross section between neutral atoms is nearly independent
of the collision energy, so that the mean free path $\lambda$ does not depend
on $v$. In plasmas, where Coulomb interaction is dominant, the mean free path
increases with $v^2$. Thus in the further calculations we set
$\lambda=\lambda _0 (v/v_0)^{\alpha}$ with $\alpha =0$ for neutral gas and
$\alpha =2$ for a fully ionised plasma.

Introducing eqs.(\ref{gM}) to (\ref{fz}) into eq.(\ref{j}), we get for the
mass and energy flux
\begin{equation}
\label{jM}
j_{\rm M}=2 \beta^{\alpha}
\left(x \beta \frac{dn}{dz} +
 x(2x^2-3) n\frac{d\beta }{dz}-\frac{2bn}{\beta }\frac{1}{x}\right)h(x)dx
\end{equation}
\begin{equation}
\label{jE}
j_{\rm E}=\beta^{\alpha +2}\!\!\left( x^3 \beta \frac{dn}{dz} +x^3 (2x^2\!-3)
n\frac{d\beta }{dz}\!-\!\frac{2bn}{\beta } x \right ) h(x)dx
\end{equation}
with $x=v/\beta$ and $h(x)=2\sqrt{\pi}m \Delta F
(\lambda_0/v_0^{\alpha})x^{2+\alpha} e^{-x^2}$.
\\Integrating these equations over all x, we obtain the conditions for zero
mass and energy flux:
\begin{equation}
\label{FOM}
\beta\frac{dn}{dz} F_{3+\alpha} +n\frac{d
\beta}{dz} [2 F_{5+\alpha} -3 F_{3+\alpha} ]-\frac{2bn}{\beta} F_{1+\alpha} =0
\end{equation}
\begin{equation}
\label{FOE}
\beta\frac{dn}{dz} F_{5+\alpha} +n\frac{d
\beta}{dz} [2 F_{7+\alpha} -3 F_{5+\alpha} ]-\frac{2bn}{\beta} F_{3+\alpha} =0
\end{equation}

The abbreviation $F_k$ stands for the integral $F_k=\int_0^{\infty} x^k
e^{-x^2} dx$. The different arguments of $F_k$ in eqs.(\ref{FOM}) and
(\ref{FOE}) can be eliminated by the recurrence formula $F_{k+2}=(k+1)/2
\times F_k$, so that we
finally obtain the equilibrium conditions for mass and energy transport:
\begin{equation}
\label{FM}
\frac{2+\alpha}{2} \beta \frac{dn}{dz}+ \frac{(2+\alpha)(
1+\alpha )}{2} n \frac{d \beta}{dz}-\frac{2bn}{\beta}=0
\end{equation}
\begin{equation}
\label{FE}
\frac{4+\alpha}{2} \beta \frac{dn}{dz}+ \frac{(4+\alpha)( 3+\alpha )}{2} n
\frac{d \beta}{dz}-\frac{2bn}{\beta}=0
\end{equation}

It is immediately evident that this is a set of linear equations for the
gradients $d \beta /dz $ and $dn/dz$. If there are no external forces
$(b=0)$, there exists only the trivial solution $d\beta /dz=0$ and $dn/dz=0$.

In the presence of volume forces such as gravitation eqs.(\ref{FM}) and
(\ref{FE}) are inhomogeneous and allow a non-trivial solution, which
constitutes a fixed relation between temperature gradient and particle
density gradient or, because of quasi-neutrality, also between temperature
and mass density gradient. Eliminating $b$ leads to the relation
\begin{equation}
\beta \frac{dn}{dz}+(2\alpha+5) n \frac{d \beta }{dz}=0
\end{equation}
with the solution $ n \beta^{(2\alpha+5)}= \rm{const.}$ Replacing the number
density by the mass density $\rho$ and $\beta$ by $\sqrt{2kT/m}$ we finally
find the condition $\rho\, T^{5/2}= \rm{const.}$ for neutral gas and $\rho\,
T^{9/2}= \rm{const.}$ for a fully ionised plasma.

This result is independent from the actual magnitude of the coefficient of
thermal conduction, if it is influenced by magnetic fields or not. The only
conditions, which have to be satisfied, are the existence of local thermal
equilibrium and the condition, that transport of kinetic energy of particles
is the dominant transfer mechanism.

\section{Observations}
During the last years several spatially resolved x-ray emission measurements
of galaxy clusters have become available, especially from the Chandra and
XMM-Newton satellites. Though temperature determinations are still rather
poor due to projection effects and necessary background corrections,
intensity profiles and from these density profiles have been obtained with
high accuracy. But in spite of the remaining error margin the classical
cooling flow model can definitely be ruled out as an explanation of the
observed data (see Kaastra et al. \cite{kaastra}). Neither the radial
temperature profiles fit to those expected from this model nor do the spectra
show the expected intensity of metal lines of lower ionisation levels.

\begin{figure}[htbp]
  \centering
  \includegraphics[width=\columnwidth,angle=0]{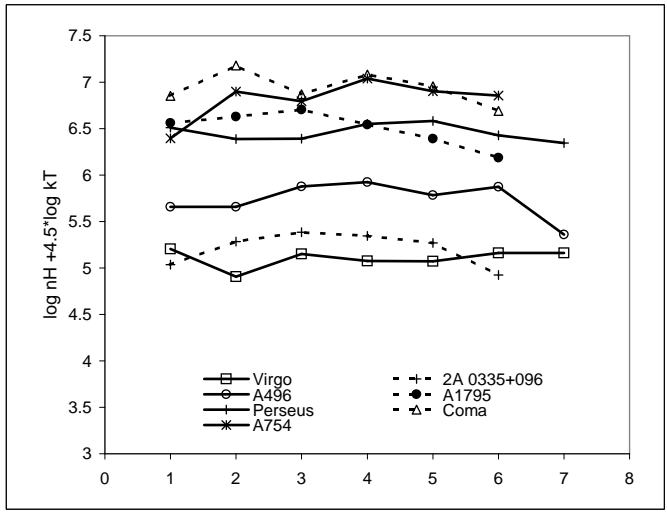}
  \caption{\textit{$kT*\rho^{2/9}$ for consecutive annular channels starting
   from cluster center for the 7 most luminous clusters given by Kaastra et
   al. \cite{kaastra}}}
  \label{kaas}
\end{figure}

Good agreement with observations can, however be obtained with a model based
on thermal conduction with inclusion of the Dufour effect. We have applied
the model to the data given by Kaastra et al. \cite{kaastra}. In
fig.\ref{kaas} the quantity $T\times \rho ^{2/9}$ is plotted for the seven
most luminous clusters, for which data are given by Kaastra et al. Only the
clusters with the highest X-ray flux ($>5\times 10^{-11}
\rm{erg\:cm^{-2}sec^{-1}}$, 0.1-2.5 keV) have been used, as for these clusters
the temperature profiles appear sufficiently accurate. The data points in
fig.\ref{kaas} refer to temperatures in consecutive annular channels starting
from centre to the boundary of the cooling region. (For details see Kaastra
\cite{kaastra}). For all seven clusters the constancy of $T
\times\rho ^{2/9}$ over the cooling region is evident. This can be regarded
as a confirmation that thermal conduction with inclusion of the Dufour effect
is the dominant energy transfer mechanism and that radiative cooling is less
important.

\begin{figure}[htbp]
  \centering
  \includegraphics[width=\columnwidth,angle=0]{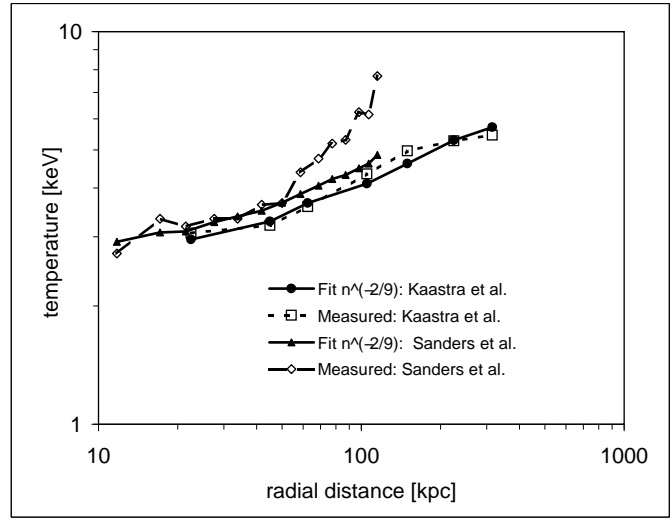}
  \caption{\textit{Temperature profile of the Perseus cluster.
  Comparison of measured temperature with profiles derived from
  measured density by $\rho\,T^{9/2}$=const., fitted at the cluster centre.}}
  \label{pers}
\end{figure}

Data from Perseus cluster have also been obtained by Chandra (see Sanders et
al. \cite{sanders}). Fig.\ref{pers} shows a comparison of the temperature
profiles obtained from direct measurements by Chandra and XMM-Newton compared
with data derived from the corresponding density profiles, applying $\rho
T^{9/2}=\rm{const.}$ The curves are fitted at the cluster centre. As can be
seen, the curves derived from density data agree very well with each other
and with the direct measurements from XMM-Newton, while the temperatures from
Chandra deviate considerably in the outer region. It should be mentioned,
however, that the Chandra data are obtained by averaging angular resolved
profiles which exhibit a large spread ($>1.5$ keV) in the outer region.

Whether radiative loss can completely neglected or has at least to be added
as a correction to the temperature profile resulting from pure heat
conduction, cannot be decided from the present data. For the model this would
require a better knowledge of the absolute magnitude of the thermal
conductivity. This requires an accurate theory of the reduction effects by
magnetic fields and more knowledge on the distribution of these fields over
the cluster volume.

\section{Conclusions}
We have shown that thermal conduction appears to be the dominant energy
transport mechanism and that a pure conduction model can describe the
observed temperature profiles, when energy flux induced by density gradients
is added to the balance. The relative temperature gradients are small
compared to the associated density gradients. But temperature changes by a
factor of two or three, as they have been found in the central regions of the
clusters mentioned above and which are reported also for other "cooling flow"
clusters by several authors (see e.g. Voigt et al. \cite{fab}), can well be
understood from the Dufour effect, as the gas density may change by a few
orders of magnitude in the vicinity of large galaxies. It appears likely that
most of the observed temperature drop in the "cooling flows" around these
galaxies is not caused by radiative cooling but by the "conductive cooling"
associated with the density gradients.

It should be noted that the Dufour effect is not restricted to the cores of
galaxies, but it should be important also in the hot corona of individual
stars, where it can help to explain the observed temperature gradient from
the surface into the interstellar medium.

Applying the model to complete galaxy clusters leads to the result that the
plasma temperature should continuously rise with increasing distance from the
centre. Observational data are contradictory in the outer region. Some
authors claim a temperature drop at the outskirts of the observable region,
others find constant or increasing temperature (see Mushotzky \cite{mush} and
references given there). Most of the recent papers find increasing
temperatures at least out to the point, where the increase of the error bars
exceedes the expected temperature gradients.

Also the conduction model should be taken with care at the low densities in
the outer regions, as the conditions of local thermal equilibrium are no
longer satisfied due to the extremely large mean free path of electrons. But
it can be stated that, if thermal conductivity were close to the Spitzer
value, clusters would have cooled off in the time since formation, unless the
clusters are embedded in a very hot but thin intergalactic plasma. The
existence of a diffuse high energy x-ray background would be a strong hint to
the existence of such an intergalactic plasma.

Measurements of the background in the energy range above that of the
intracluster gas are rare. Observations from the HEAO-1 satellite (Boldt
\cite{boldt}) have found a continuous background, which can be interpreted as
bremsstrahlung in the temperature range of 40 to 50 keV, but the spatial
resolution of these early data is so poor that it cannot be decided, if it is
really radiation from a continuously distributed plasma or from the
superposition of individual point sources. As a plausible explanation for the
formation of a hot continuous background plasma one could think of the
relativistic matter jets expelled by quasars. During the time period of
several Gyr this matter should have become homogeneously distributed in
space.

It has been shown that the conduction model presented in this paper can
explain some observations in the central regions of the intracluster plasma,
but it deserves more confirmation by detailed measurements, until the
astrophysics community will be convinced. Also the extension to the outer
regions of galaxy clusters would bring additional confirmation, but this
requires the extension of spatially resolved measurements into the high
energy range, as they are planned with the launch of future satellites like
Astro-E2, which are scheduled for the next years.

\end{document}